\documentclass[twocolumn,superscriptaddress,floatfix,preprintnumbers,nofootinbib,prd,10pt]{revtex4-2}
\pdfoutput=1
\usepackage[colorlinks=true,breaklinks=true]{hyperref}
\usepackage[normalem]{ulem}
\usepackage[utf8]{inputenc}
\hypersetup{allcolors=[rgb]{0.0 0.0 0.6},linkcolor=[rgb]{0.75 0.05 0.05}}
\usepackage{amsmath,amssymb}
\usepackage{epsfig}  
\usepackage{graphicx}
\usepackage{appendix}

\usepackage{epsfig}  
\usepackage{graphicx}   
\usepackage{slashed}       
\usepackage{tikz}
\usepackage{url}
\usepackage{color}
\usepackage{multirow}
\usepackage{comment}
\usepackage{amssymb}
\usepackage[capitalize]{cleveref}
\usepackage{bm}
\usepackage{soul}

\DeclareMathOperator{\GeV}{GeV}

\DeclareMathOperator{\MeV}{MeV}

\DeclareMathOperator{\gs}{g_{*,s}}

\newcommand{\beq}{\begin{equation}}
\newcommand{\eeq}{\end{equation}}

\newcommand{\df}{\Delta f}
\newcommand{\feq}{f^{\mathrm{eq}}}
\newcommand{\iin}{\mathrm{in}}
\newcommand{\tin}{t_\iin}
\newcommand{\Neff}{N_\mathrm{eff}}
\newcommand{\DNeff}{\Delta N_\mathrm{eff}}
\newcommand{\tG}{\widetilde{\Gamma}}

\newcommand{\TRH}{T_\mathrm{RH}}

\allowdisplaybreaks

\bibpunct{[}{]}{,}{n}{}{,}

\begin{document}

\title{Strong cosmological constraints on the neutrino magnetic moment}

\author{Pierluca Carenza}\email{pierluca.carenza@fysik.su.se}
\affiliation{The Oskar Klein Centre, Department of Physics, Stockholm University, Stockholm 106 91, Sweden
}

\author{Giuseppe Lucente}
\email{giuseppe.lucente@ba.infn.it}
\affiliation{Dipartimento Interateneo di Fisica “Michelangelo Merlin”, Via Amendola 173, 70126 Bari, Italy}
\affiliation{Istituto Nazionale di Fisica Nucleare -- Sezione di Bari, Via Orabona 4, 70126 Bari, Italy}
\affiliation{Kirchhoff-Institut f\"ur Physik, Universit\"at Heidelberg, Im Neuenheimer Feld 227, 69120 Heidelberg, Germany}
\affiliation{Institut f\"ur Theoretische Physik, Universit\"at Heidelberg, Philosophenweg 16, 69120 Heidelberg, Germany}

\author{Martina Gerbino}\email{gerbinom@fe.infn.it}
\affiliation{Istituto Nazionale di Fisica Nucleare, Sezione di Ferrara, via Giuseppe Saragat 1, I-44122
Ferrara, Italy}

\author{Maurizio Giannotti}
\email{mgiannotti@unizar.es}
\affiliation{Centro de Astropart{\'i}culas y F{\'i}sica de Altas Energ{\'i}as (CAPA), Universidad de Zaragoza, Zaragoza, 50009, Spain}
\affiliation{Department of Chemistry and Physics, Barry University, 11300 NE 2nd Ave., Miami Shores, FL 33161, USA}

\author{Massimiliano Lattanzi}
\email{lattanzi@fe.infn.it}
\affiliation{Istituto Nazionale di Fisica Nucleare, Sezione di Ferrara, via Giuseppe Saragat 1, I-44122
Ferrara, Italy}

\smallskip

\begin{abstract}
A sizable magnetic moment for neutrinos would be evidence of exotic physics.
In the early Universe, left-handed neutrinos with a magnetic moment would interact with electromagnetic fields in the primordial plasma, flipping their helicity and producing a population of right-handed (RH) neutrinos.
In this work, we present a new calculation of the production rate of RH neutrinos in a multi-component primordial plasma and quantify their contribution to the total energy density of relativistic species at early times, stressing the implications of the dependence on the initial time for production. We find that current cosmological data exclude values of the magnetic moment $\mu \gtrsim 1.6\times 10^{-11}~\mu_B$, while future cosmological experiments will be able to probe non-thermal production of RH neutrinos, becoming competitive with stellar limits.

\end{abstract}
\maketitle

\section{Introduction}
\label{sec:introduction}
The existence of a sizable neutrino magnetic moment (NMM) $\mu$ is a fascinating possibility, which has generated a vivid interest in experimental and phenomenological communities in the recent years. 
If equipped with (transition) magnetic moments, neutrinos would couple to photons through the effective Lagrangian terms
\begin{equation} 
	L=-\frac12 \mu^{ij} \, \overline \psi_i \sigma_{\alpha\beta}\psi_j\, F^{\alpha\beta} \,,
 \label{eq:lagrangian}
\end{equation}
where $\psi$ is the neutrino field, $F$ the electromagnetic field tensor, $\alpha$, $\beta$ are Lorentz indices, and $i$, $j$ are the flavor indices. 
The first experimental constraint goes back to the very neutrino discovery experiment by Cowan and Reines~\cite{Cowan:1957pp}.
Analyzing electron recoil spectra in (anti)-neutrino electron scattering, the authors derived the bound $\mu\lesssim 10^{-9}\,\mu_B$, where $\mu_B=e/2m_e$ denotes the Bohr magneton.
Later searches gave considerably stronger 
constraints~\cite{Reines:1976pv,TEXONO:2002pra,MUNU:2003peb,Derbin:1993wy,TEXONO:2010tnr,Beda:2012zz,Borexino:2017fbd,XENONCollaboration:2022kmb}.
The current experimental sensitivity is slightly below \mbox{$10^{-11}$ $\mu_{B}$}. 
Although neutrinos with a Dirac mass are expected to have a magnetic moment~\cite{Fujikawa:1980yx}, this is predicted to be quite small, $\mu\approx 3\times 10^{-19} \times (m_\nu/1\,{\rm eV}) \,\mu_B$.
Thus, a sizable magnetic moment would be evidence of exotic physics. 
Furthermore, even though several extensions of the standard model predict a larger neutrino magnetic moment, 
the predictions change substantially for Dirac and Majorana neutrinos, with the latter allowed more naturally to have a large $\mu$~\cite{Barbieri:1987xm,Bell:2005kz,Davidson:2005cs,Bell:2006wi}.
Hence, a discovery of a sufficiently large magnetic moment would also shed light on the nature of the neutrino field. 

A non-vanishing neutrino magnetic moment would have profound astrophysical and cosmological consequences.
Stellar evolution provides some of the strongest bounds on $\mu$.
The resulting tree-level neutrino-photon interaction contributes substantially to several cooling mechanisms in stars (see, e.g., Ref.~\cite{Heger:2008er} for a comprehensive discussion), most notably to the plasmon decay $\gamma\to\bar \nu \nu$.
A recent analysis of the cooling of red giant stars~\cite{Capozzi:2020cbu} provided the constraint $\mu\lesssim1.2\times10^{-12}\mu_{B}$, about a factor of 5 below the current experimental sensitivity.\footnote{Different observables could be sensitive to different combinations of the elements of the magnetic moment matrix $\mu^{ij}$.  For example, terrestrial experiments using reactor sources (electron anti-neutrinos) are sensitive to the $\mu^{ei}$ components.
On the other hand, bounds based on energy loss in stars, such as the one in Ref.~\cite{Capozzi:2020cbu}, are sensitive to all of the combinations of $\mu^{ij}$.
}

Furthermore, for Dirac neutrinos, a non-vanishing magnetic moment would induce the production of right-handed (RH) neutrinos via spin-flip of left-handed (LH) neutrinos in an external electromagnetic field. 
For a supernova, where LH neutrinos are trapped and RH neutrinos are not, this would imply a very efficient energy loss mechanism~\cite{Barbieri:1988nh,Barbieri:1988xw,Notzold:1988kz}. 
Spin-flip processes may also play a significant role in the early Universe, given the large number of charged particles generating electromagnetic fields required for this process to take place. The production of RH neutrinos in this case would affect the predictions of the light-elements abundances because of the increased number of relativistic degrees of freedom $N_\mathrm{eff}$~\cite{Morgan:1981psa,Morgan:1981zy,Fukugita:1987uy,Loeb:1989dr}.   
In the past, the above observables were used to derive the constraint $\mu\lesssim6.2\times10^{-11}\mu_{B}$~\cite{Elmfors:1997tt},
a result considerably less stringent than the current astrophysical limits.
The analysis of Ref.~\cite{Elmfors:1997tt}, however, did not account for spin-flip processes in the electromagnetic field generated by charged particles other than electrons and positrons. 
In the early Universe, especially above the phase transition of the Quantum Chromodynamics (QCD), we expect a large number of charged particles and thus more efficient spin-flip processes. 
Current and upcoming cosmological surveys severely constrain $\Neff$, leading to tighter bounds on $\mu$. 
A recent analysis in this direction was performed in Ref.~\cite{Li:2022dkc}. The study presented a detailed calculation of the neutrino chirality flipping rate in a thermal plasma, including a careful and consistent consideration of soft scattering and the plasmon effect in finite temperature field theories. This analysis found that a NMM above $2.7\times10^{-12}\mu_B$ is excluded by current CMB (Cosmic Microwave Background) and BBN (Big-Bang Nucleosynthesis) measurements of $\Delta N_{eff}$. Here, we revisit and improve on this calculation of RH neutrino production in the early Universe, extensively discussing the cosmological implications. 

In Sec.~\ref{sec:rate}, we discuss the thermal field theory methods employed in the calculation of the rate. In Sec.~\ref{sec:cosmological}, we discuss the cosmological implications of a sizable NMM. Moreover, we show that the requirement of not exceeding the number of effective relativistic species allowed by cosmological observations results in a strong constraint on $\mu$.  Finally, in Sec.~\ref{sec:conclusions}, we conclude by discussing incoming cosmological experiments able to probe the freeze-in regime of RH neutrino production. Two Appendices follow.

\section{RH neutrino production rate}
\label{sec:rate}
Neutrinos with a magnetic moment interact with the surrounding plasma through current-current interactions with the electromagnetic fields generated by charged particles. Information about the medium and its electromagnetic fluctuations is encoded in the imaginary part of the photon polarization tensor $\Pi_{\mu\nu}$, obtained at one loop by summing over loops with all the charged particles in a plasma.
Ref.~\cite{Elmfors:1997tt} calculated the net right-handed (RH) neutrinos production rate $\Gamma$ in a hot plasma using the thermal field theory approach
\begin{equation}
\begin{split}
\Gamma=&\frac{\mu^{2}}{2\pi}\int_{0}^{\infty}dk\,k\int_{-\infty}^{\infty}dk_{0}\theta\left(-K^{2}(K^{2}+4pk_{0}+4p^{2})\right)   \,\\
&\epsilon(k_{0})\frac{K^{4}}{k^{2}}\Bigg[\frac{\epsilon(k_{0})}{e^{\frac{|p_{0}+k_{0}|}{T}}+1}+\frac{\epsilon(p_{0}+k_{0})}{e^{\frac{|k_{0}|}{T}}-1}+\\
&+\epsilon(p_{0}+k_{0})\theta(-k_{0})-\epsilon(k_{0})\theta(-k_{0}-p_{0})\Bigg]\\
&\left[\left(1+\frac{k_{0}}{p}+\frac{K^{2}}{4p^{2}}\right)\mathcal{A}_{T}(K)-\left(1+\frac{k_{0}}{2p}\right)^{2}\mathcal{A}_{L}(K)\right]\,,
   \label{eq:rate}
\end{split}
\end{equation}
where $(p_{0},p)$ is the neutrino four-momentum, $K=(k_{0},k)$ the photon four-momentum, $T$ the temperature of the thermal bath, $\theta$ the Heaviside theta function, $\epsilon$ the sign function, and the sources of electromagnetic fields are encoded in the photon spectral functions $\mathcal{A}_{T,L}$, defined as
\begin{equation}
    \mathcal{A}_{T,L}(K)=-\frac{1}{\pi}\frac{{\rm Im}\Pi_{T,L}}{|K^{2}-{\rm Re}\Pi_{T,L}|^{2}+|{\rm Im}\Pi_{T,L}|^{2}}\,,
\label{eq:atl}
\end{equation}
in terms of the transverse $\Pi_T$ and longitudinal $\Pi_L$ components of the photon polarization tensor. In a single component plasma with a fermion of mass $m_f$ and charge $e_f$ the photon polarization tensor is given by~\cite{Weldon:1982aq}
\beq
\begin{split}
\Pi_{L}(m_f,e_f)&=-\frac{K^{2}}{k^{2}}u^{\mu}u^{\nu}\Pi_{\mu\nu}\,,\\
\Pi_{T}(m_f,e_f)&=-\frac{1}{2}\Pi_{L}+\frac{1}{2}g^{\mu\nu}\Pi_{\mu\nu}\,,\\
\end{split}
\eeq
where $g^{\mu\nu}$ is the metric tensor, $u^\mu$ the plasma four-velocity, which in the rest frame reduces to \mbox{$u^{\mu} = (1, 0, 0, 0)$}, while 
\beq
\begin{split}
&u^{\mu}u^{\nu}\Pi_{\mu\nu}=e_f^{2}\int \frac{dq \,q^{2}}{2E_{q}\pi^{2}}\frac{1}{e^{E_{q}/T}+1}\,\\
&\bigg[2-\frac{(2E_q+k_0)^2-k^2}{4q k}\ln\left(\frac{k_0^{2}-k^{2}+2(E_{q}k_0+q k)}{k_0^{2}-k^{2}+2(E_{q}k_0-q k)}\right) \\
& +\frac{(2E_q-k_0)^2-k^2}{4q k}\ln\left(\frac{k_0^{2}-k^{2}-2(E_{q}k_0+q k)}{k_0^{2}-k^{2}-2(E_{q}k_0-q k)}\right)\bigg]\,,
\end{split}
\label{eq:uupi}
\eeq
and
\beq
\begin{split}
&g^{\mu\nu}\Pi_{\mu\nu}=2e_f^{2}\int \frac{dq\,q^{2}}{2E_{q}\pi^{2}}\frac{1}{e^{E_{q}/T}+1}\\
&\bigg[2+\frac{k_0^{2}-k^{2}+2m_{f}^{2}}{4q k}\ln\left(\frac{k_0^{2}-k^{2}+2(E_{q}k_0-q k)}{k_0^{2}-k^{2}+2(E_{q}k_0+q k)}\right)\\
&-\frac{k_0^{2}-k^{2}+2m_{f}^{2}}{4q k}\ln\left(\frac{k_0^{2}-k^{2}-2(E_{q}k_0-q k)}{k_0^{2}-k^{2}-2(E_{q}k_0+q k)}\right)\bigg]\,,\\
\label{eq:gpi}
\end{split}
\eeq
with $E_q =\sqrt{q^{2}+m_{f}^{2}}$ the energy of the fermion in the loop. Note that this formalism includes all the effects proportional to $\mu^{2}$, as plasmon decay and electron-positron annihilation in neutrino-antineutrino pairs, spin-flip transitions and Cherenkov processes~\cite{Elmfors:1997tt}.
Among these, plasmon decay is negligible in the Early Universe~\cite{Li:2022dkc}.\\
In a multi-component plasma, the full polarization tensor is obtained by summing  Eq.~\eqref{eq:uupi} and Eq.~\eqref{eq:gpi} over all the species $f$:     \begin{equation}
    \Pi_{T,L}=\sum_{f}\Pi_{T,L}(m_{f},e_{f})\,.
\end{equation}

\begin{figure}
    \centering
    \includegraphics[width=0.49\textwidth]{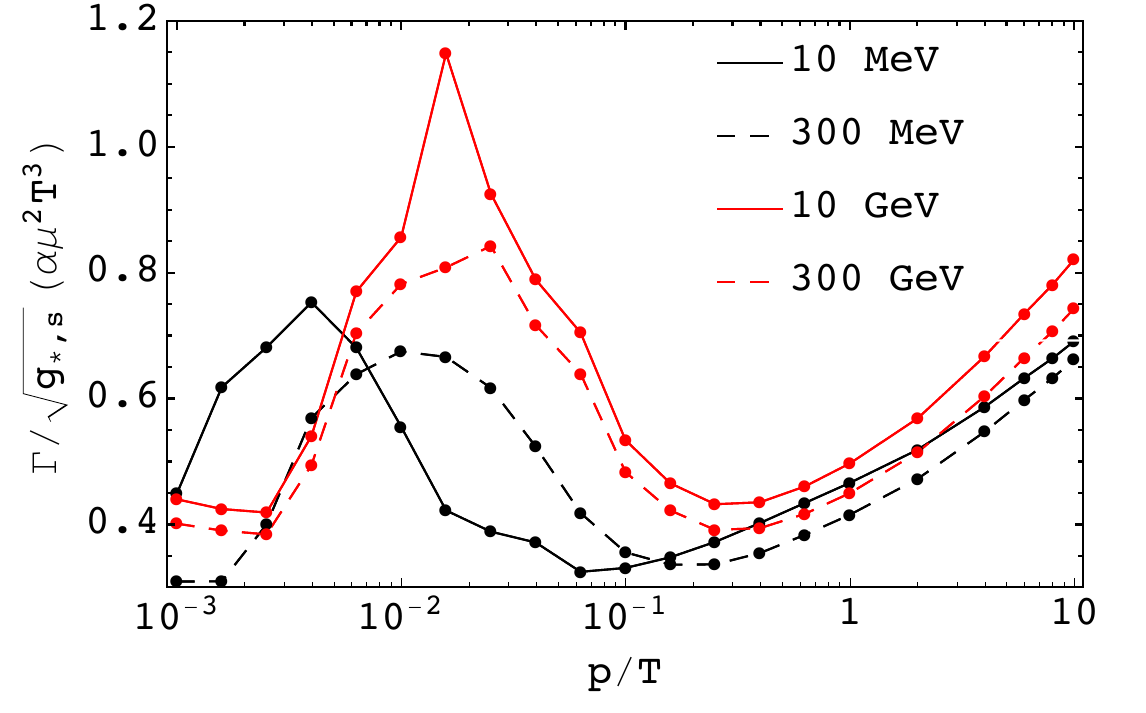}
        \includegraphics[width=0.49\textwidth]{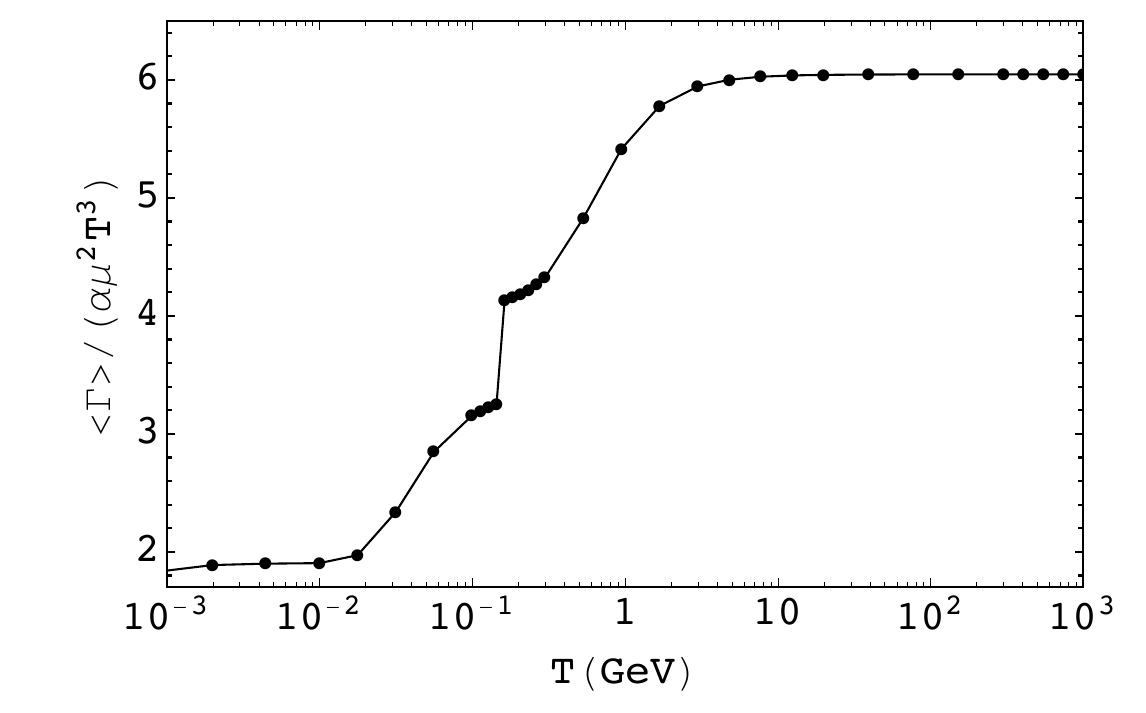}
    \caption{\emph{Upper panel:} Production rate $\Gamma$ for RH neutrinos as a function of the neutrino momentum in terms of $p/T$, for different values of temperature. The lines are a linear interpolation between the values of $p/T$ at which the rate is computed, marked by circles. \emph{Lower panel:} Thermally-averaged production rate $\langle \Gamma \rangle$ as a function of temperature. The solid curve shows a linear interpolation between the values of $T$ at which the rate is computed, marked by circles.} 
    \label{fig:rates}
\end{figure}

Our study extends the previous results~\cite{Elmfors:1997tt}, valid for a mono-component plasma of massless electrons, to the more realistic case of a multi-component primeval plasma, including finite mass effects. 
We underline that the numerical integration of the production rate in Eq.~\eqref{eq:rate} is stiff when the photon approaches the on-shell condition. 
In this limit, the imaginary part of the polarization tensor vanishes, ${\rm Im}\Pi_{T,L}\to0$, and Eq.~\eqref{eq:atl} reduces to a Dirac delta function $\delta(K^2 - {\rm Re}\Pi_{T,L})$.
Therefore, we can replace the integration over the photon energy with just the contribution coming from the on-shell photon. 
This is at the origin of the peak in the production rate $\Gamma$ at low values of the neutrino momentum ($p/T\ll 1$) shown in the upper panel of Fig.~\ref{fig:rates}. However, this affects only slightly our results, which depend mostly on large momenta. 
As expected, the rate increases with the number of charged species in the plasma, as encoded by the effective number of entropy degrees of freedom $g_{*,s}(T)$. 
We find that $\alpha^{-1} \mu^{-2} T^{-3}\Gamma(p/T)/ \sqrt{g_{*,s}(T)}$ is approximately constant for $T\gtrsim10$~MeV.

In the lower panel of Fig.~\ref{fig:rates} we show the full shape of $\langle\Gamma\rangle/(\alpha\mu^2\,T^3)$, obtained by averaging over a thermal neutrino distribution. 
The series of drops below $T\approx100\,\mathrm{GeV}$ are due to changes in $g_{*,s}(T)$.
Our calculation smoothly interpolates between different temperatures, leading to $\langle\Gamma\rangle = 1.91\,\alpha\mu^2 T^3\,$, where the brackets denote a thermal average over a plasma containing only electrons at $T= 10$~MeV, and $\langle\Gamma\rangle = 6.04\,\alpha\mu^{2}T^{3}$ at $T= 100$~GeV. This result can be compared to~\cite{Li:2022dkc}, which evaluated the RH neutrino production rate using a different approach, obtaining $\langle\Gamma\rangle = 6.47\,\alpha\mu^{2}T^{3}$.

\section{Cosmological implications} 
\label{sec:cosmological}
The phase-space distribution function for RH neutrinos 
$f^R(t,q)$ of comoving momentum $q=a p$, with $a$ the cosmological scale factor, evolves according to the Boltzmann equation~\cite{Elmfors:1997tt}:
\begin{equation}
    \partial_t f^R = - \Gamma \left(f^R- \feq \right), \label{eq:boltz}
\end{equation}
where $\feq(t, q)$ is the (local) equilibrium distribution function. 
We take $\feq$ to be the distribution function of the left-handed neutrinos $f_\nu$, which are kept in thermal equilibrium with the primordial plasma at temperature $T$ by the weak interactions
%%%%%%%%
\begin{equation}
\label{eq:f_nu_f_eq}
\feq \simeq f_\nu = \frac{1}{\exp{\left(\frac{q}{aT}\right)}+1} \,.
\end{equation}
The evolution equation is solved for $f^R$ over a grid of values of the comoving momentum, together with the initial condition $f^R_{in}=0$, i.e., no initial population of RH neutrinos. 

The initial integration time, corresponding to a temperature $T_\iin$, can be chosen freely and may affect the outcome. The integration always ends at $T_\mathrm{fin} = 10 \,\MeV$ as, at lower temperatures, the spin-flip rate becomes rapidly negligible and no more RH neutrinos are produced.

The ratio $\left\langle\Gamma\right\rangle/H$, $H$ being the Hubble parameter, scales as $\mu^2 T$ in the radiation-dominated era, for constant effective number of entropy degrees of freedom $g_{*,s}$.
Thus, the production of RH neutrinos is dominated by the high-temperature regime, and our results might depend on the initial time chosen for the integration. 
Defining the decoupling temperature $T_d $ through $\left\langle\Gamma\right\rangle/H |_{T_d} =1$ , RH neutrinos quickly thermalize roughly at the initial time, if $T_\iin\gg T_d$, and their final abundance will be independent of $T_\iin$. In this case, the abundance relative to active neutrinos depends only on entropy injections in the plasma happening at $T\lesssim T_d$.
Note that since $\left\langle\Gamma\right\rangle/H \propto T$, it takes some time for decoupling to complete, and significant entropy injection to the RH component can take place even at $T<T_d$.
On the other hand, if $T_\iin \lesssim T_d$, the abundance of RH neutrinos depends on $T_\iin$, since their production happens out-of-equilibrium, in a ``freeze-in'' kind of process, with $T_\iin$ setting the time interval during which RH neutrinos can be efficiently produced. Both regimes are correctly captured by solving the Boltzmann equation, as discussed in Appendix~\ref{app:freezein}. The extra contribution~\footnote{In the standard cosmological model, the number of relativistic degrees of freedom beyond photons is given by active neutrinos only. The predicted value is $N_\mathrm{eff}=3.044$ \cite{Mangano:2001iu,deSalas:2016ztq,Akita:2020szl,Bennett:2020zkv,Froustey:2021azz}, and we define $\DNeff\equiv N_\mathrm{eff} - 3.044$.} 
$\Delta N_\mathrm{eff}$ to  $N_\mathrm{eff}$ due to the population of a single species of RH neutrinos at a given time is obtained as
\begin{equation}
    \Delta N_\mathrm{eff}(T) = \frac{\int \mathrm{d}p\, p^3 f^R(p,T)}{\int \mathrm{d}p \,p^3 f_\nu(p,T)}\,.
    \label{eq:DNeff}
\end{equation}

\begin{figure}[t!]
    \centering
    \includegraphics[width=0.9\linewidth]{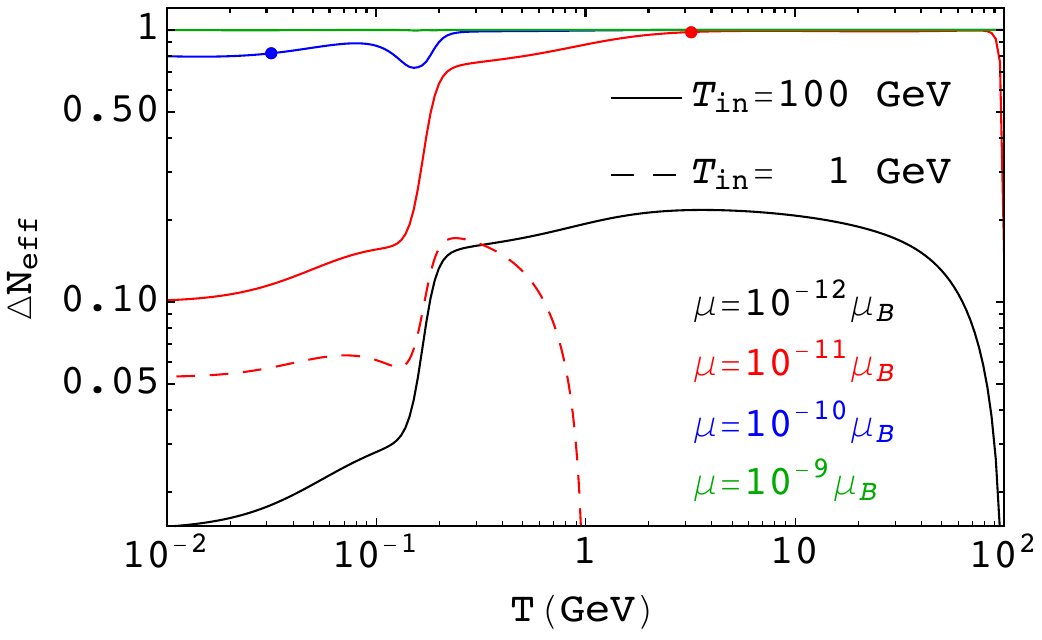}
    \caption{Evolution of $\Delta N_\mathrm{eff}$ due to the population of a single species of RH neutrinos, as per Eq.~\eqref{eq:DNeff}. 
    The solid lines are obtained for $T_\mathrm{in}=100\,\mathrm{GeV}$ and different values of $\mu$, while the dashed line for $\mu=10^{-11}~\mu_B$ and $T_\mathrm{in}=1\,\mathrm{GeV}$.  
    The dots mark the values of the decoupling temperature $T_d$.}
    \label{fig:Neff_vs_T}
\end{figure}

In Fig.~\ref{fig:Neff_vs_T}, we show the evolution of $\Delta N_\mathrm{eff}(T)$ for different values of $\mu$ setting the initial temperature to ${T_\mathrm{in}=100\,\mathrm{GeV}}$. On each curve,
we mark with a small circle the corresponding decoupling temperature. There are no circles on the curves for $\mu=10^{-12}\,\mu_B$ and $\mu=10^{-9}\,\mu_B$, as those lie outside the temperature range shown in the plot (at higher and lower temperatures, respectively).
For the largest values of $\mu$ in the plot, $\Delta N_\mathrm{eff}$ quickly reaches a saturation limit corresponding to a thermal abundance at the photon temperature. 
The drop in $\Delta N_\mathrm{eff}(T)$ at later times is due to entropy injection in the plasma happening after the decoupling of RH neutrinos (particularly evident is the drop at the QCD phase transition, $T_{\rm QCD}\approx 150 \,\MeV$). In this regime, the final abundance is thus basically set by the decoupling temperature. 

For lower values of $\mu$ (\mbox{$\mu \lesssim 10^{-11}~\mu_B$} for \mbox{$T_\mathrm{in}=100\,\GeV$}), RH neutrinos do not have enough time to thermalize. In this regime $\DNeff$ decreases with decreasing $\mu$ due both to entropy production and  a smaller RH population.
As expected, $\Delta N_\mathrm{eff}$ becomes negligibly small for vanishing $\mu$. In the plot we also include the case for $\mu=10^{-11}\,\mu_B$ and $T_\mathrm{in}=1\,\mathrm{GeV}$, shown as a dashed red line. 
The lower initial temperature implies that, despite the rapid initial growth, $\DNeff$ remains significantly smaller than the  contribution obtained for the same value of $\mu$,  but with $T_\mathrm{in}=100\,\mathrm{GeV}$ (solid red line). 

In the following, we take the ``late-time'' value of $\DNeff(T)$ as the one relevant for cosmological observations. $\DNeff(T)$ is constant for $T\lesssim10\,\MeV$ since no significant production of RH neutrinos takes place after that time, and the comoving density of active neutrinos does not change in the limit of instantaneous decoupling.\footnote{Neutrino decoupling is not instantaneous, implying that neutrinos actually benefit from a small fraction of entropy release from electron-positron annihilation. Neglecting this effect does not change significantly our results.}
We then use $\DNeff(T=10\,\MeV)$ as the late-time value of $\DNeff$.

\section{Constraints on the NMM}
\label{sec:constraints}

\begin{figure}[t!]
    \centering
\includegraphics[width=0.9\linewidth]{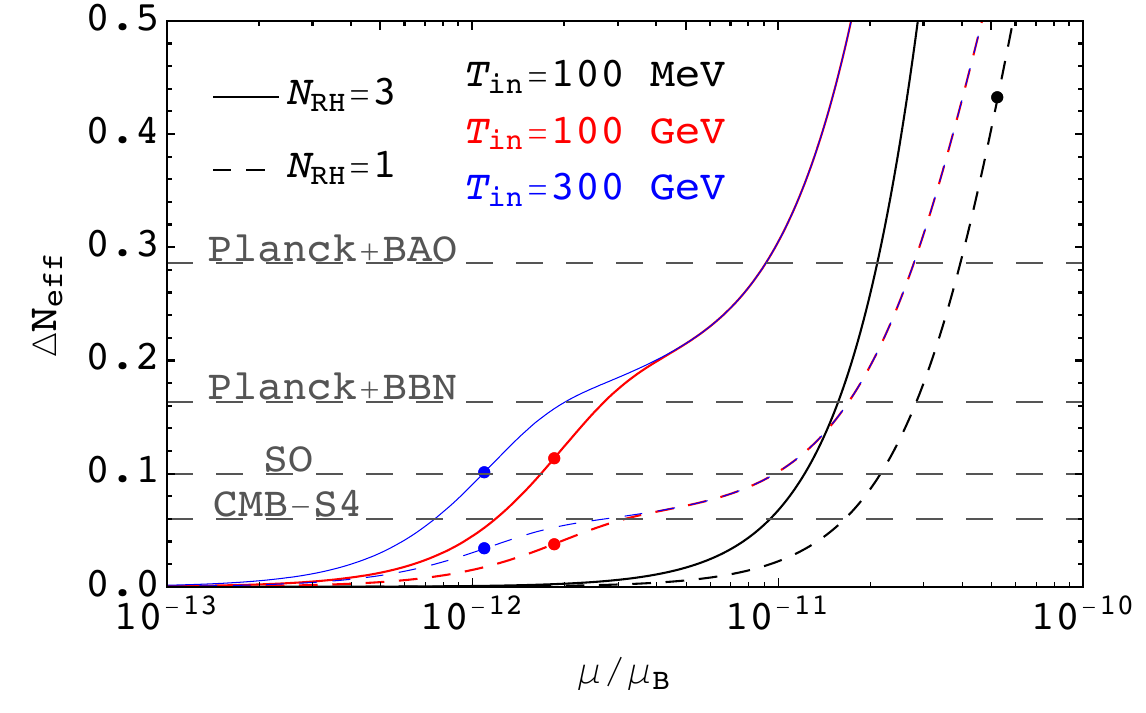}
    \caption{The late-time value of $\Delta N_\mathrm{eff}$ due to the population of three (solid) or one (dashed) species of RH neutrinos as function of $\mu$ for different initial temperatures, zoomed in the region where $\Delta\,N_{\rm eff}<0.5$. The horizontal dashed lines indicate the current 95\% bound (Planck+BAO and Planck+BBN) and the $2\sigma$ sensitivity of future experiments (SO and CMB-S4). The dots mark the values of $\mu$ giving 
    $\left\langle \Gamma \right\rangle / H = 1 $ at $T=T_\iin$.}
    \label{fig:Neff_vs_mu}
\end{figure}

In Fig.~\ref{fig:Neff_vs_mu} we report the late-time value of $\Delta N_\mathrm{eff}$ as a function of $\mu$ for different initial temperatures (${T_{\rm in}=100~\MeV}$ in black, $T_{\rm in}=100~\GeV$ in red, ${T_{\rm in}=300~\GeV}$ in blue), in the case of one (dashed lines) or three (solid lines) species of RH neutrinos.
We can identify three main regions, depending on the size of $\mu$.
As expected, in the limit of very large $\mu$, the RH population remains coupled to the active neutrino species until late times, hence providing a large contribution to $N_\mathrm{eff}$. 
For lower values of $\mu$, RH neutrinos are still in thermal equilibrium with the plasma at early times, but the decoupling happens earlier. 
In this scenario, RH neutrinos are efficiently diluted by entropy production and their contribution to $N_\mathrm{eff}$ is smaller.
For even smaller values of $\mu$, RH neutrinos are never in thermal equilibrium with the active species. 
A residual non-thermal population can then develop and contribute very feebly, yet non-vanishingly, to $N_\mathrm{eff}$.
As further discussed in Appendix~\ref{app:approximations}, the results that we obtain for $\Neff$ in both the limits of large and small $\mu$ are different from the ones obtained in Ref.~\cite{Li:2022dkc}, since we do not employ the instantaneous decoupling approximation, nor we assume a priori that RH neutrinos are in thermal equilibrium with the primeval plasma.
Precisely, since $\Gamma/H$ scales linearly with $T$, sizable entropy transfer from the SM plasma to the RH neutrinos can take place for quite some time after the decoupling, implying that the approximation of instantaneous decoupling is particularly crude. Thus, neglecting entropy production happening after decoupling can severely underestimate the value of $\Delta N_\mathrm{eff}$ associated to a given value of the NMM.  \\
Concerning the assumption of thermal equilibrium, one might argue that since $\Gamma/H$ increases at high temperatures, the condition for equilibrium $\Gamma/H \gg 1$ can always be met at some very early time. This argument however requires to be addressed more carefully. 
There are several reasons why an extrapolation to very high energies might be inappropriate. First, from a strictly observational point of view, we have no information about the cosmic thermal history above the BBN energy scale. Second, new physics might play a role above the electroweak scale,  altering the thermal evolution of the plasma. 
Finally, our effective description of the neutrino-photon coupling induced by the NMM is expected to break down at some high energy scale, that depends on the physics inducing the NMM. 
Thus, it is not granted that one can always take $T_\mathrm{in}$ large enough for thermal equilibrium to be established. 
Solving the Boltzmann equation with vanishing initial RH population, we are able to describe the scenario in which thermal equilibrium is established as well as the scenario in which the interactions are not strong enough to generate a thermal population.
In Fig.~\ref{fig:Neff_vs_mu} we mark with small circles the values of $\mu$ that yield $\left\langle \Gamma \right\rangle / H = 1 $ at the initial temperature.
To the right of the circles, $\left\langle \Gamma \right\rangle / H > 1 $ at $T_\iin$, corresponding to the thermal regime. 
All curves overlap in the right part, where the final abundance of RH neutrinos is independent of $T_\iin$ since $T_d\ll T_\iin$. 
Conversely, the region to the left of the circles traces the freeze-in regime, $\left\langle \Gamma \right\rangle / H < 1 $ at $T_\iin$.

With our approach we are able to smoothly compute the value of $\Delta N_{\rm eff}$ as a function of the NMM and compare it with observations to set constraints. The horizontal dashed lines in Fig.~\ref{fig:Neff_vs_mu}  represent the $95\%$ bayesian credible upper bounds on $\Delta N_\mathrm{eff}$ from a combination of current cosmological data (observations of CMB anisotropies from Planck, combined with Baryon Acoustic Oscillations, BAO, from a compilation of large-scale-structure surveys~\cite{Planck:2018vyg}, and Big-Bang Nucleosythesis data, BBN~\cite{Fields:2019pfx})\,\footnote{The current 95\% credible interval for $\Neff$ from Planck+BAO is $\Neff = 2.99^{+0.34}_{-0.33}$~\cite{Planck:2018vyg} and for Planck+BBN is $\Neff =2.83\pm0.38$~\cite{Fields:2019pfx}.} and the expected $2\sigma$ sensitivity on $\Delta N_\mathrm{eff}$ from future CMB surveys (Simons Observatory, SO, \cite{SimonsObservatory:2018koc} and CMB-S4 \cite{Abazajian:2019eic}).
Assuming $T_{\rm in}=100~\MeV$ (black line in Fig.~\ref{fig:Neff_vs_mu}) current measurements probe the thermal regime and imply:
\begin{equation}
\begin{split}
   & \mu < 2.1 \times 10^{-11} \mu_B \quad (\textrm{Planck+BAO})\,, \\
   &  \mu < 1.6\times 10^{-11} \mu_B \quad (\textrm{Planck+BBN})\,. 
\end{split}
    \label{eq:muboundMeV}
\end{equation}
However, the value of $T_\mathrm{in}$ is somewhat arbitrary and can affect the final abundance. 
Values of $T_{\rm in} > 100\,\MeV$ can be safely assumed, provided that other particles species heavier than electrons and positrons are included in the plasma. Therefore, it is necessary to follow the RH neutrino thermalization across a large temperature range, from $T_{\rm in}$ to today, by properly taking into account the charged particle species contributing to the RH neutrino production. At this regard, the formalism discussed in Sec.~\ref{sec:rate} and the use of the Boltzman equation allow us to accurately reconstruct the evolution of $\Delta\,N_{\rm eff}$ due to the population of RH neutrinos produced starting from ${T_{\rm in} > 10~\MeV}$. This represents the main improvement of this work compared to previous literature~\cite{Elmfors:1997tt}.  Taking $T_\mathrm{in}=100\,\mathrm{GeV}$ (red line) as our benchmark, current bounds become:
\begin{equation}
\begin{split}
   & \mu < 9.1 \times 10^{-12} \mu_B \quad (\textrm{Planck+BAO})\,, \\
   &  \mu < 2.6\times 10^{-12} \mu_B \quad (\textrm{Planck+BBN})\,. 
\end{split}
    \label{eq:muboundGeV}
\end{equation}
Future cosmological data will instead go deep into the freeze-in regime, for three RH neutrino families, and test values as low as $\mu = 1.7 \times 10^{-12} \mu_B$ (SO) and $\mu = 1.2 \times 10^{-12} \mu_B$ (CMB-S4). The formalism developed here allows us to constrain even such low values of the neutrino magnetic moment, which do not lead to the establishment of a thermal population of RH neutrinos. A detailed comparison with the results obtained in the previous literature can be found in Appendix~\ref{app:approximations}.

\section{Discussion and conclusions}
\label{sec:conclusions}
In this work, we have revisited the calculation of the RH neutrino production in the early Universe, extensively discussing the cosmological implications and computing current bounds and sensitivities of future cosmological surveys on the neutrino magnetic moment $\mu$. As mentioned in Sec.~\ref{sec:constraints}, one key aspect of the cosmological analysis presented here is that the choice of $T_\iin$ is somehow arbitrary: it can be as large as the highest reheating temperature $T_\mathrm{RH}$ currently allowed by the non-observation of
primordial $B$-modes in the CMB. 
Assuming a perfectly efficient reheating and instantaneous thermalization of standard model particles after inflation, $T_\iin \sim T_\mathrm{RH}\sim V^{1/4}< 1.6\times 10^{16}\,\GeV$ \cite{Planck:2018jri}, $V$ being the energy density at inflation. 

In the thermal regime, the final RH abundance only depends on the value of $g_{*,s}$ at decoupling, and increasing $T_\iin$ above $100\,\GeV$ does not change the Planck+BAO bound on $\mu$, while it slightly affects the Planck+BBN bound, assuming no other species beyond the particle content of the standard model, compare the red and blue lines in Fig.~\ref{fig:Neff_vs_mu}. The freeze-in abundance is instead proportional to $\mu^2 T_\iin$ for constant $g_{*,s}$, as shown in Appendix~\ref{app:freezein}.
The future sensitivity on $\mu$ will then increase for $T_\iin > 100\,\GeV$, scaling as $1/\sqrt{T_\iin}$.\,\footnote{We have explicitly checked that when deriving constraints on $\mu$ using the rates computed at $10^3\,\GeV$, we find $\mu< 4.3\times10^{-13}~\mu_B$ for CMB-S4, in line with the expected $\sqrt{10}$ improvement.}
Note, however, that the effective interaction Lagrangian in Eq.~\eqref{eq:lagrangian} will break down above some energy threshold determined by the physics governing the magnetic moment, and  other  assumptions made in our calculation could fail for $T_\iin \gg 100\,\GeV$. 
For example, several theories beyond the standard model predict new degrees of freedom that could contribute to $g_{*,s}$ at those temperatures.
Furthermore, the spin-flip rate in Eq.~\eqref{eq:rate} assumes thermal equilibrium of the standard model plasma. 
Though possible (see, e.g., Refs.~\cite{Davidson:2000er,Harigaya:2013vwa,Mukaida:2022bbo}), a quick,  effectively instantaneous, thermalization of the standard model plasma after reheating is not guaranteed, as it depends on the details of the inflation model. A careful extension of the formalism laid down in this work is therefore required to project the sensitivity on $\mu$ in the regime $T_\iin\gg 100\,\mathrm{GeV}$.

Moreover, it cannot be excluded that the Universe underwent other reheating episodes beyond the one associated to the end of inflation. In fact, the latest of such reheating events can happen at a temperature $\TRH$ as low as a few MeV without contradicting available observations~\cite{deSalas:2015glj}. The final abundance is reduced for $T_\iin \sim \TRH \le T_d$, resulting in looser constraints on $\mu$; see e.g., the $T_\iin = 100\,\MeV$ curve in Fig.~\ref{fig:Neff_vs_mu}.
Our choice of $T_\iin = 100\,\GeV$ amounts to assume a standard thermal history at $T\le 100\,\GeV$ and that the physics  responsible for the generation of neutrino magnetic moment lies at or above the electroweak scale.

\begin{figure}[t!]
    \centering
\includegraphics[width=0.9\linewidth]{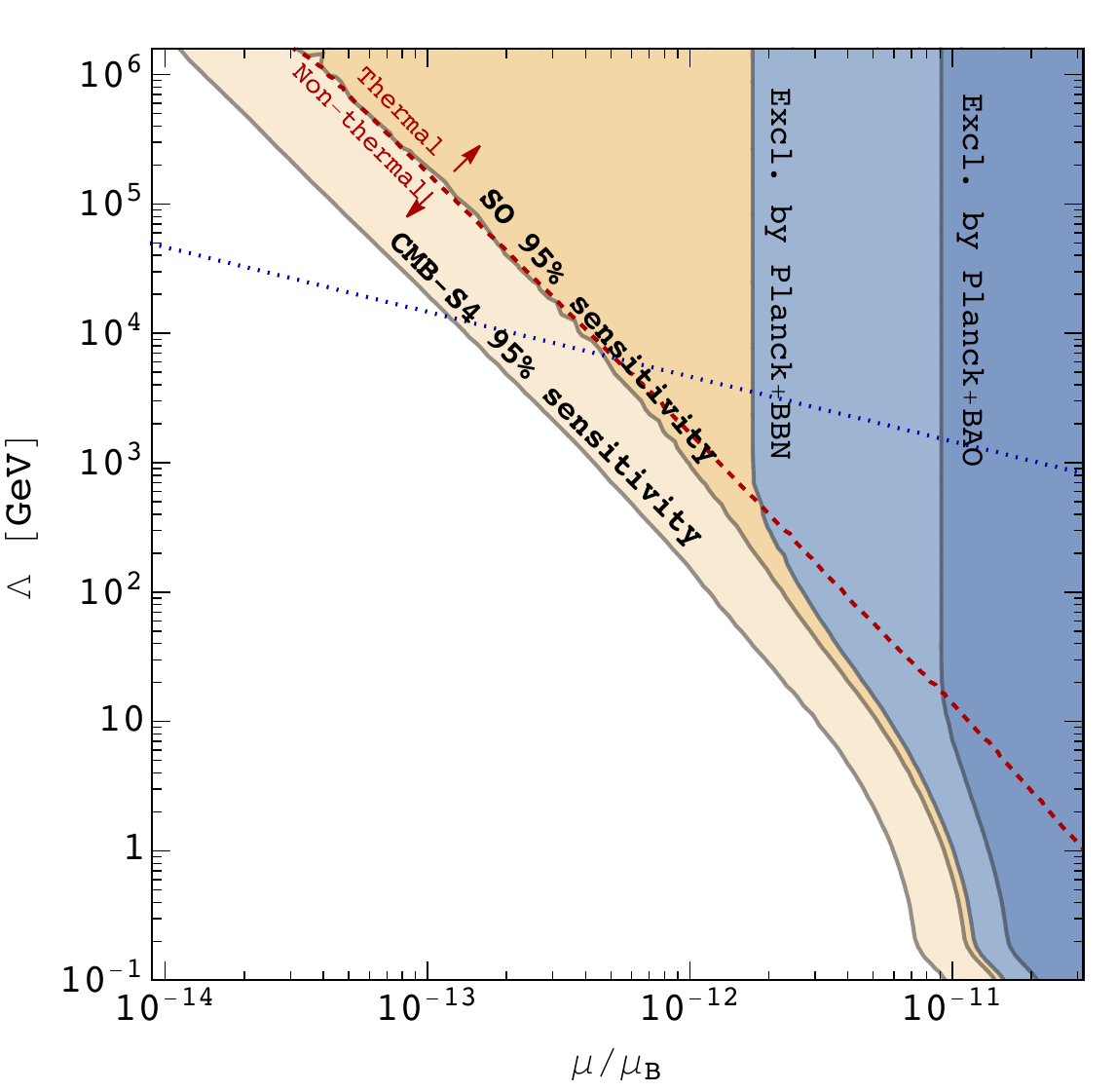}
    \caption{Current constraints (blue) and discovery potential of future experiments (orange) in the $(\Lambda,\,\mu)$ plane, where $\Lambda$ is the energy scale at which production of RH neutrinos begins. The red dashed line separates the regions of thermal and non-thermal production. The yield of RH neutrinos is practically independent of $\Lambda$ above this line. The blue dotted line shows the prediction $\mu \sim m_X/(16 \pi^2 \Lambda^2)$ for the magnetic moment generated by new physics at the scale $\Lambda$, mediated by a charged heavy fermion with mass $m_X=1\,\GeV$ \cite{Xu:2019dxe}.}
    \label{fig:mu_T_2D}
\end{figure}

In order to better understand the role of the initial temperature, in Fig.~\ref{fig:mu_T_2D} we plot current constraints and discovery potentials for $\mu$ assuming that production of RH neutrinos begins at a arbitrary energy scale $\Lambda\equiv T_\iin$. There are two main ways to interpret this plot. If one assumes that the physics generating the magnetic moment lies well above the energy scale associated to reheating, then $\Lambda$ can be identified with $T_\mathrm{RH}$. Otherwise, $\Lambda$ can be identified with the scale of the new physics itself. Note that in this case the constraints shown in Fig.~\ref{fig:mu_T_2D} are conservative, since we are neglecting any production that might have occurred at $T>\Lambda$. The red dashed line in the figure separates the regions of thermal and non-thermal production,\,\footnote{In more detail, the red dashed line corresponds, for each $\mu$, to the initial temperature that gives a yield equal, within 1\%, to the yield that would be obtained in the limit $T_\iin\to\infty$ (i.e., for $T_\iin \gg T_d$).} and clearly shows how next generation experiments will probe the non-thermal regime.
In Fig.~\ref{fig:mu_T_2D} the blue regions are probed by current data (Planck+BAO, Planck+BBN). It is possible to note that these exclusion regions are independent of $\Lambda$ when crossing the dashed red line, denoting the condition of equilibrium. Reasonably, from the particle physics point of view, we expect $\Lambda\gtrsim 10^{3}$~GeV, while cosmological observations allow for a non-standard thermal history for $\Lambda\gtrsim 10$~MeV. In this context, future cosmological probes such as SO and CMB-S4, whose sensitivities are represented by the orange regions, will be capable of fully exploring the non-thermal production of RH neutrinos, shedding light on the possible new physics above $\Lambda$. 

We remark that the constraints discussed here are irreducible, since they are obtained neglecting additional non-standard interactions between neutrinos and other, perhaps exotic, particles. Indeed, these interactions, required to justify a neutrino magnetic moment larger than the standard model prediction, $\mu \gg \mathcal{O}(10^{-19}\mu_{B})$, might contribute to the RH neutrino production in the early Universe, increasing the value of $\Delta N_{\rm eff}$. 

\begin{acknowledgments}

We thank Edoardo Vitagliano and Damiano Fiorillo for helpful comments on the draft, and Patrick Stengel for useful feedback on the thermalization after reheating. We warmly thank Shao-Ping Li and Xun-Jie Xu for discussion and comparison with their recent work~\cite{Li:2022dkc}.\\
This article/publication is based upon work from COST Action COSMIC WISPers CA21106, supported by COST (European Cooperation in Science and Technology). We are grateful to Alessandro Mirizzi for his contributions  in the initial stages of the project. The work of P.C. is supported by the European Research Council under Grant No.~742104 and by the Swedish Research Council (VR) under grants  2018-03641 and 2019-02337.
GL is supported by the European Union’s Horizon 2020 Europe research and innovation programme under the Marie Skłodowska-Curie grant agreement No 860881-HIDDeN.
MG and ML acknowledge support from the COSMOS network (www.cosmosnet.it) through the ASI (Italian Space Agency) Grants no.\ 2016-24-H.0, 2016-24-H.1-2018, and 2019-9-HH.0. We acknowledge the use of CINECA HPC resources from the InDark project in the framework of the INFN-CINECA agreement.
\end{acknowledgments}

\appendix

\section{Details on the freeze-in regime}
\label{app:freezein}
In the freeze-in regime, it is possible to derive the explicit dependence of $\DNeff$ on $\mu$ and the initial temperature $T_\mathrm{in}$ and far from entropy production events, when the equilibrium distribution does not depend on time, i.e. $\feq(q,t) = \feq(q)$ .
Introducing the deviation from equilibrium $\df(q,t) \equiv f^R(q,t) - \feq(q)$, it is straightforward to write a formal solution of the Boltzmann equation
\begin{equation}
\begin{split}
    \df(q,t)&= -\feq(q) \exp\left[-\int_{\tin}^t \Gamma\, dt'\right] =\\
    &    = -\feq(q) \exp\left[-\int_{z}^{z_\iin}\frac{\Gamma}{H(1+z')} dz' \right], \label{eq:solf}
    \end{split}
\end{equation}
where we use $q=ap$ as the momentum variable to ensure that $\partial_t \feq = 0$, and the initial condition $\df(q,\, \tin) = -\feq(q)$ because we assume a vanishing initial population of RH neutrinos. In the following, we will use the subscript ``$\iin$'' to denote quantities evaluated at the initial time.
We now write the RH neutrino production rate as
\begin{equation}
    \Gamma = \alpha \mu^2 \sqrt{g_{*,s}(T)} T^3 \tG(y, T) \,,
\end{equation}
where the momentum variable is $y=p/T = q/aT$. Note that $\tG$ does not depend on the temperature far from entropy production events, and its temperature dependence is anyway very mild, even when accounting for entropy production (see the lower panel in Fig.~\ref{fig:rates}).

Using the expression for the Hubble rate in the radiation-dominated era, we can recast the integral in Eq.~\eqref{eq:solf} as
\begin{equation}
\begin{split}
    \int_{z}^{z_\iin}\frac{\Gamma}{H(1+z')} dz'
     &= \sqrt{\frac{45}{4\pi^3}}\alpha \mu^2 m_\mathrm{pl} \int_{z}^{z_\iin} \frac{\tG\,T}{1+z}  dz \simeq \\
     &\simeq     \sqrt{\frac{45}{4\pi^3}}\alpha \mu^2 m_\mathrm{pl} \left.\left(\tG \,T \right)\right|_{z_\iin} = \left.\frac{\Gamma}{H}\right|_{z_\iin} \, ,
     \label{eq:intappr}
     \end{split}
\end{equation}
where the approximate equality holds for $z\ll z_\iin$, and when the initial time is far from from entropy production events. Precisely, since $T/(1+z) \propto \gs(T)$ and $\tG$ itself is weakly dependent on temperature, the integrand varies significantly only in proximity of entropy production events, and the value of the integral is dominated by the high-temperature contribution. Note the somehow fortunate cancellation between the factors of $\sqrt{g_{*,s}}$ in $\Gamma$ and $H$,  since the effective number of degrees of freedom for energy density $g_{*,r}= g_{*,s}$ for $T\gg 1\,\MeV$.

Since we are focusing on the freeze-in regime, $\Gamma/H \ll 1$. From the above discussion, the requirement that this inequality holds at $T_\iin$ guarantees that it also holds at $z<z_\iin$, and that its integrated value over $d\ln(1+z)$ is itself much smaller than $1$. In terms of $\mu$ and $T_\iin$, the condition $\Gamma_\iin/H_\iin \ll 1$ reads:
\begin{equation}
0.47 \times \left(\frac{\mu}{10^{-12} \mu_b}\right)^2 
\left(\frac{T_\iin}{100\,\GeV} \right) \tG_\iin \ll 1 \,.
\label{eq:freezein_cond}
\end{equation}
\\
We can thus expand the exponential in Eq.~\eqref{eq:solf} to first order in its argument, and then compute $\Delta \Neff$. This yields, for a single species of RH neutrinos:
\begin{align}
    \DNeff(T\ll T_\iin) = \frac{120}{7\pi^4}\left(\frac{\gs(T)}{\gs(T_\iin)}\right)^{4/3}    \int_0^\infty dy \,y^3 \frac{\Gamma_\iin}{H_\iin} \feq,
    \label{eq:Dneffappr}
\end{align}
where $\Gamma_\iin = \Gamma(y, T_\iin)$.
The temperature dependence of the right-hand side is encoded in the $\gs^{4/3}$ factor, and can be understood as follows. In the freeze-in regime, the production of RH neutrinos mostly happens at high temperatures, thus the RH abundance quickly saturates shortly after the initial time and stays constant for $T \ll T_\iin$. The energy density of RH neutrinos relative to the active ones can only change when more active neutrinos are produced due to entropy injection in the plasma.
Substituting the explicit expression of $\Gamma_\iin/H_\iin$ with the one shown in Eq.~\eqref{eq:intappr}, we obtain
\begin{equation}
\begin{split}
    \DNeff=&\frac{120}{7\pi^4}\sqrt{\frac{45}{4\pi^3}} \alpha \mu^2 T_\iin m_\mathrm{pl} \\
    &\quad\left(\frac{\gs(T)}{\gs(T_\iin)}\right)^{4/3}\int_0^\infty dy \,y^3 \tG_\iin f^\mathrm{eq} \, .
\end{split}
\end{equation}

This quantity depends on the initial temperature through the explicit factor $T_\iin \gs(T_\iin)^{-4/3}$, and also through the integrand. However, when $\gs$ is constant, this dependence reduces to the only factor $T_\iin$. In this regime, ${\DNeff \propto \mu^2 T_\iin}$. Thus, a given observational bound on $\DNeff$ translates, in the freeze-in regime defined by Eq.~\eqref{eq:freezein_cond}, to a bound on $\mu^2 T_\iin$. In other words, constraints on $\mu$ in the freeze-in regime scale as $T_\iin^{-1/2}$. 
Note that, since $\Gamma/H \propto \mu^2 T$, the decoupling temperature  scales as $\mu^{-2}$. 
This ensures that, for constant $\gs$, the rescaled constraint will still correspond to the freeze-in regime. To illustrate this, let us  consider an initial temperature in the freeze-in regime, $T_\iin \ll T_d $, and the associated constraint $\mu < \overline{\mu}$. Rescaling $T_\iin$ to $T'_\iin > T_\iin$ would produce a new upper bound (assuming freeze-in) $\overline{\mu}' = \overline{\mu} \sqrt{\frac{T_\iin}{T'_\iin}}$. Since the decoupling temperature $T'_d$ associated to $\overline{\mu}'$ is $T'_d = T_d \frac{\overline{\mu}^2}{\overline{\mu}'^2}$, we find that $T'_\iin/T'_d = T_\iin/T_d \ll 1$.

\section{Impact of the different cosmological analyses}
\label{app:approximations}

\begin{figure}[t!]
    \centering
\includegraphics[width=0.9\linewidth]{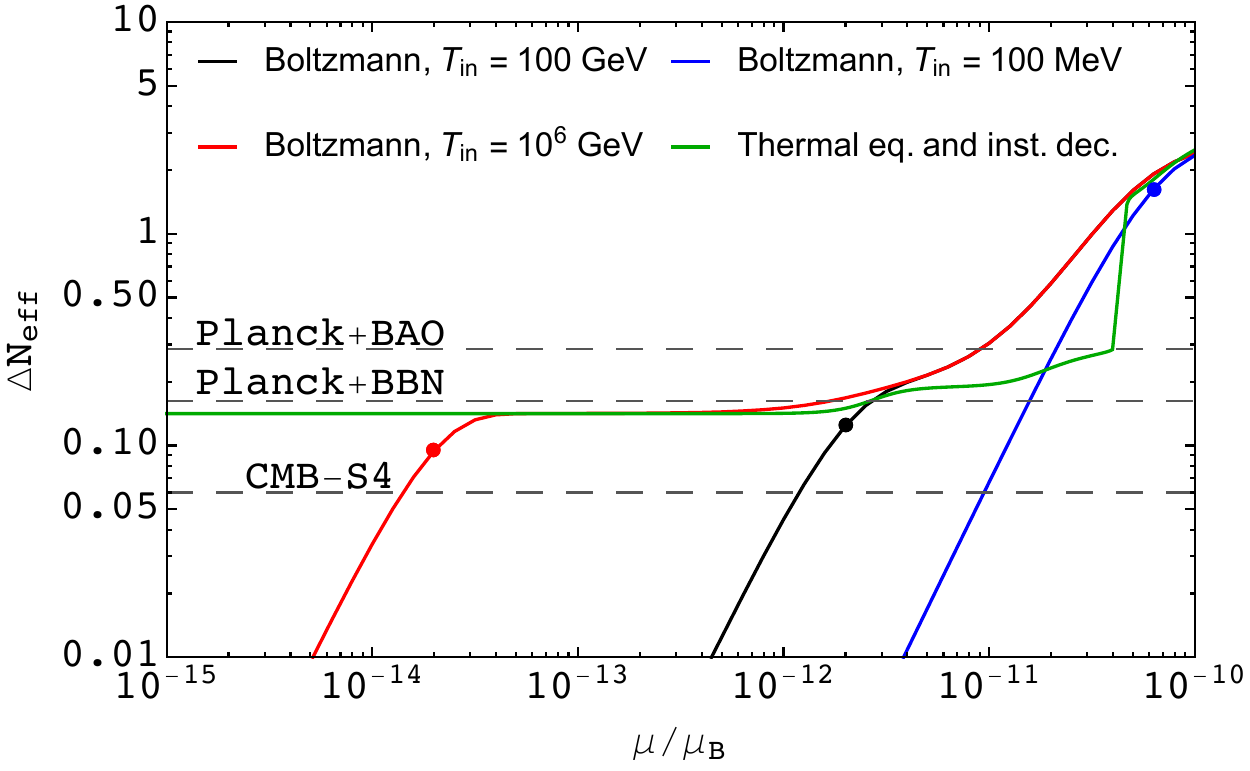}
    \caption{Late-time value of $\Delta N_\mathrm{eff}$ due to the population of three species of RH neutrinos as function of $\mu$ for different initial temperatures and calculation methods (see text for details). The horizontal dashed lines indicate the current 95\% bound (Planck+BAO and Planck+BBN) and the $2\sigma$ sensitivity of future experiments (CMB-S4). The dots mark the values of $\mu$ giving 
    thermal equilibrium at the initial temperature $T=T_\iin$.}
    \label{fig:comparison}
\end{figure}

Here we discuss modeling assumptions and approximations that enter the computation of $\Neff$ and subsequent bounds on the NMM. 
We find it useful, for the purpose of this discussion, to show in Fig.~\ref{fig:comparison} the value of $\Neff$ for three RH neutrinos as a function of $\mu$, as obtained with different approaches. The black, red and blue curves have been obtained by direct integration of the Boltzmann equation, for three different initial temperatures, namely $T_{\rm in}=100$~GeV, $T_{\rm in}=10^6$~GeV and $T_{\rm in}=100$~MeV, respectively, and assuming a vanishing population of RH neutrinos at the initial time. 

First, we compare our results with those of Ref.~\cite{Elmfors:1997tt}, where the primeval plasma is composed only by electrons and positrons, restricting the analysis to temperatures $T\le m_\mu \simeq 100\,\MeV$. A bound $\mu\lesssim 6.2\times 10^{-11}\mu_B\,$ was derived by requiring that RH neutrinos have already decoupled from the thermal bath at the highest temperature considered in the analysis, i.~e. $\Gamma<H$ at $T=100~\MeV$, with $\Gamma$ the thermally-averaged interaction rate. Since a species decoupling after $\mu^+\mu^-$ annihilation is at a temperature equal or larger than the active neutrinos, this amounts to requiring that each RH neutrino species contributes with $\DNeff<1$ to the radiation density. Applying the same criterion ($\Gamma<H$ at 100 MeV) using our own calculation of the rate, we obtain $\mu < 5.3\times10^{-11}~\mu_B$, as represented by the blue dot in Fig.~\ref{fig:comparison}, roughly in good agreement with Ref.~\cite{Elmfors:1997tt}.

Current observations constrain $\Delta\,N_{\rm eff} < 0.286$ (Planck + BAO), leading to a stronger bound on $\mu$.  Indeed, ignoring any production that might have happened at epochs earlier than $T\simeq100~\MeV$, as implicitly done in Ref.~\cite{Elmfors:1997tt}, would yield $\mu \lesssim 2.1\times 10^{-11}\mu_B$, as shown by the blue curve in Fig.~\ref{fig:comparison}.
However, the value of $T_\mathrm{in}$ is somehow arbitrary and can affect the final abundance. Values of $T_{\rm in} > 100\,\MeV$ can be considered, provided that other particles species in the plasma are included. This is the main improvement of this work. For $T_{\rm in}=100~\GeV$ (the black curve in Fig.~\ref{fig:comparison}) we obtain the bound 
\begin{equation}
    \mu\lesssim 9.1\times 10^{-12}~\mu_B\,.
\end{equation}
Notice that this result corresponds to full thermalization at $T\gtrsim 4\,\GeV$. Therefore, this is insensitive to further increases of the initial temperature above 100~GeV, as confirmed by the $T_{\rm in}=10^{6}~\GeV$ case (i.e., by noting that the black and red curves in Fig.~\ref{fig:comparison} overlap in the region constrained by the Planck+BAO dataset).

The dataset combination Planck + BBN prefers smaller values of $\Neff$ than Planck+BAO, albeit with a slightly large uncertainty, and as a result provides the stronger constraint $\Delta N_{\rm eff}\lesssim 0.163$, leading to $\mu<2.7\times 10^{-12}~\mu_{B}$ for $T_\mathrm{in} = 100\,\GeV$. In this case, the largest allowed value of $\mu$ correspond to thermal equilibrium at $T\gtrsim 50\,\GeV$, which is fairly close to $T_\mathrm{in}$. We can thus expect that the bound will saturate for a somehow larger choice of the initial temperature. This can be concluded by noting that the black and red curves in Fig.~\ref{fig:comparison} start to diverge for values of the NMM close to the upper bound. Indeed, we have explicitly verified that the bound saturates at $\mu<1.7\times 10^{-12}~\mu_{B}$ for large enough values of $T_\mathrm{in}$, as shown by the red line.

We now comment on the difference between the green line and the other cases previously discussed. In our analysis, we follow the decoupling of the RH neutrinos by solving the relevant Boltzmann equation with vanishing initial population, and for different values of the initial temperature $T_\mathrm{in}$. 
This yields the black, red and blue lines in Fig.~\ref{fig:comparison} for $T_{\rm in}=100$~GeV, $T_{\rm in}=10^{6}$~GeV and $T_{\rm in}=100$~MeV, respectively, as detailed before. A different approach, used e.g. in Ref.~\cite{Li:2022dkc}, would consist in assuming that a thermal population is established at early times and that decoupling happens instantaneously at the time when $\Gamma/H=1$. One should then use entropy conservation to find the ratio between the RH and photon temperatures at late times and compute the corresponding value of $\DNeff$. This yields the green line in Fig.~\ref{fig:comparison}.

The first difference between the two approaches is related to the instantaneous decoupling approximation, as evident by comparing the black and green lines in Fig.~\ref{fig:comparison}. Before commenting further on the origin of this difference, we discuss the effects on the NMM bounds.
The initial temperature for the red curve, $T_\mathrm{in} = 10^6 \,\GeV$, has been chosen to correspond to thermal equilibrium for $\mu \gtrsim 2\times 10^{-14}\mu_B$, as represented by the red dot, so that the differences between the two curves at larger values of the NMM are due to the treatment of decoupling. It can be clearly seen in Fig.~\ref{fig:comparison} that the instantaneous decoupling can significantly underestimate the value of $\DNeff$, and thus lead to (artificially) looser constraints on $\mu$. For Planck+BAO, the constraint would go from $\mu < 9.1\times 10^{-12}\mu_B$ to $\mu < 3.7 \times 10^{-11}\mu_B$. For Planck+BBN, the constraint would go from $\mu < 1.7\times 10^{-12}\mu_B$ to $\mu < 2.6 \times 10^{-12}\mu_B$. The latter value matches very closely the bound quoted in Ref.~\cite{Li:2022dkc}, $\mu < 2.7 \times 10^{-12}\mu_B$, as expected given that it has been obtained under the same assumptions. This bound is also very close to the one we found using the full Boltzmann approach for $T_\mathrm{in}=100\,\GeV$ (black line). This agreement should be however regarded as accidental, as it is clear that the black and green curves in Fig.~\ref{fig:comparison} have different behaviors, and only by chance cross at a value of $\DNeff$ corresponding to the Planck+BBN bound.

Given that $\Gamma/H$ scales linearly with $T$, sizable entropy transfer from the SM plasma to the RH neutrinos can take place for quite some time after the condition $\Gamma/H=1$ is realized at the decoupling temperature $T_\mathrm{dec}$, i.e. the approximation of instantaneous decoupling is particularly crude.  Thus, neglecting entropy production happening after decoupling, at $T<T_\mathrm{dec}$, can severely underestimate the value of $\Delta N_\mathrm{eff}$ associated to a given value of the NMM.  This explains the discrepancy between the black and green lines.
The solution obtained by solving the Boltzmann equation with the full temperature dependence of $\Gamma/H$ (red curve) clearly differs in the region of $10^{-12} \mu_B\lesssim\mu\lesssim 5\times10^{-11} \mu_B$. This is particularly important since this is exactly the region probed by currently available data. Note also that solving the Boltzmann equation leads to a much smoother behavior for $\DNeff(\mu)$, making the results obtained in this approach less dependent on the modeling of the expansion and interaction rates around the time of QCD decoupling.

One might naively expect that, since $\Gamma/H$ increases at large temperatures, the equilibrium condition $\Gamma/H \gg 1$ can be always met in the Early Universe. However, several factors might affect this extrapolation to  high temperatures. From an observational standpoint, there are direct probes of the cosmological history only below the BBN energy scale. Additionally, new physics might appear above the electroweak scale, potentially altering the primeval plasma. Moreover, the NMM interaction is an effective description valid up to a certain high energy threshold, depending on the physics inducing the NMM. 
Hence, it cannot be assumed that the parameter $T_\mathrm{in}$ can always be set sufficiently high to establish thermal equilibrium. By resolving the Boltzmann equation with a vanishing initial RH neutrino population, we are able to both describe the scenario in which thermal equilibrium is established as well as the scenario in which the interactions are not strong enough to generate a thermal population.

\bibliographystyle{bibi}
\bibliography{biblio.bib}

\end{document}